\documentclass[11pt]{article}
\usepackage[utf8]{inputenc}
\usepackage[left=1in,right=1in]{geometry}
\usepackage{authblk}
\usepackage[pdfstartview=FitH,pdfpagemode=None]{hyperref}
\usepackage{amsmath}
\usepackage{cite}
\title{Generalized Supergravity Equations and Generalized Fradkin-Tseytlin Counterterm}
\author[1,2]{Wolfgang M\"uck}
\affil[1]{Dipartimento di Fisica ``Ettore Pancini", Universit\`a degli Studi di Napoli ``Federico II" \authorcr Via Cintia, 80126 Napoli, Italy}
\affil[2]{Istituto Nazionale di Fisica Nucleare, Sezione di Napoli \authorcr Via Cintia, 80126 Napoli, Italy}
\affil[ ]{\texttt {mueck@na.infn.it}}
\date{}
\numberwithin{equation}{section}

\newcommand{\ie}{i.e.,\ }
\newcommand{\eg}{e.g.,\ }

\newcommand{\rmd}{\,\mathrm{d}}

\newcommand{\e}[1]{\operatorname{e}^{#1}}

\newcommand{\vev}[1]{\left\langle #1 \right\rangle}

\begin{document}
\maketitle
\begin{abstract}
The generalized Fradkin-Tseytlin counterterm for the (type I) Green-Schwarz superstring is determined for background fields satisfying the generalized supergravity equations (GSE). For this purpose, we revisit the derivation of the GSE based upon the requirement of kappa-symmetry of the superstring action. Lifting the constraint of vanishing bosonic torsion components, we are able to make contact to several different torsion constraints used in the literature. It is argued that a natural geometric interpretation of the GSE vector field that generalizes the dilaton is as the torsion vector, which can combine with the dilatino spinor into the torsion supervector. To find the counterterm, we use old results for the one-loop effective action of the heterotic sigma model. The counterterm is covariant and involves the worldsheet torsion for vanishing curvature, but cannot be constructed as a local functional in terms of the worldsheet metric. It is shown that the Weyl anomaly cancels without imposing any further constraints on the background fields. In the case of ordinary supergravity, it reduces to the Fradkin-Tseytlin counterterm modulo an additional constraint.  
\end{abstract}
\newpage

\section{Introduction}

Ten-dimensional supergravities arise in string theory as low-energy effective theories describing the dynamics of massless string excitations. The universal bosonic sector common to the type I and type II theories comprises the metric, the dilaton and the Kalb-Ramond two-form. Recently, string backgrounds have been found, which satisfy a more general set of field equations called the generalized supergravity equations (GSE), the most prominent feature of which is the absence of a scalar dilaton. 
The GSE were found in \cite{Arutyunov:2015mqj} in the context of integrable deformations of the $AdS_5\times S^5$ type II superstring sigma model \cite{Delduc:2013qra,Delduc:2014kha,Kawaguchi:2014qwa,Borsato:2016ose}, which are closely related to non-Abelian $T$-duality transformations \cite{Orlando:2016qqu,Hoare:2016wsk,Hong:2018tlp,Borsato:2018idb}.\footnote{Precursors to the GSE have appeared earlier in, e.g., \cite{Hull:1986xn,Elitzur:1994ri}.} Subsequently, they were derived from the requirement of kappa-symmetry of the Green-Schwarz (GS) sigma model in superspace \cite{Wulff:2016tju}, correcting the long-standing conjecture or conviction that on-shell supergravity is not only sufficient \cite{Howe:1983sra} but also necessary for invariance under kappa-symmetry of the GS action. In fact, the result obtained by Tseytlin and Wulff \cite{Wulff:2016tju} shows that kappa-symmetry of the GS action requires the background supergravity fields to satisfy the GSE. This resolves a related puzzle for the deformed sigma model \cite{Arutyunov:2015qva}.\footnote{A similar statement holds for BRST invariance of the classical pure spinor superstring \cite{Berkovits:2000fe} invalidating earlier claims \cite{Berkovits:2001ue} that BRST invariance impies the supergravity constraints, see \cite{Mikhailov:2012id, Mikhailov:2014qka}.} The GSE have also been studied in the context of double field theory \cite{Sakatani:2016fvh,Sakamoto:2017wor} and exceptional field theory \cite{Baguet:2016prz}. 

As mentioned above, the main difference between the GSE and ordinary supergravity is the absence of a scalar dilaton, although on-shell supergravity configurations are special solutions to the GSE. More precisely, there are two fields, a ``dilatino'' $\chi_\alpha$ and a vector $X_a$ which, in the special case of supergravity, are given by $\chi_\alpha=\nabla_\alpha \Phi$ and $X_a=\nabla_a \Phi$, respectively. These fields are common to both, the type I and type II, cases. The type II equations involve, in addition, a Killing vector $K_a$, which, combined with a Killing spinor superfield, generates a superisometry \cite{Wulff:2016tju}. 

This state of affairs raises an important question as to the consistency of string theory on such generalized supergravity backgrounds. It was argued in \cite{Arutyunov:2015mqj} that the GSE are the conditions for scale invariance of the sigma model, while Weyl invariance requires the stronger supergravity equations. This statement, however, is at odds with expectations from sigma model anomalies \cite{Wulff:2018aku} and from the fact that GSE solutions can be related by $T$-dualities to a solution of standard supergravity \cite{Hoare:2015wia}. The consistency of a sigma model is tied to the vanishing ot the Weyl anomaly, which in turn is related to the beta functions of the background fields \cite{Callan:1985ia}. For (super)strings in a (NS-NS) background satisfying the supergravity  equations, Weyl invariance is achieved by the addition of the Fradkin-Tseytlin counterterm \cite{Fradkin:1985ys} (see also \cite{Hull:1985rc, Tseytlin:1986tt, Shore:1986hk, Tseytlin:1986ws})
\begin{equation}
\label{intro:FT}
	S_{\text{FT}} = \frac1{4\pi} \int\rmd^2 \xi \sqrt{-G} R \Phi~,
\end{equation}
where $G_{IJ}$ is the worldsheet metric, $R$ its Ricci scalar, and $\Phi$ the dilaton of the background. For supergravity backgrounds with non-trivial fermionic components, the situation is a bit more subtle, because of issues connected to the quantization of the GS superstring \cite{Nissimov:1987ci,Grisaru:1988wm,Grisaru:1988jt,Grisaru:1988sa,Pasti:1988rc,Majumdar:1989sx,Bergshoeff:1989rf}. For the heterotic string, for example, it has been shown \cite{Bellucci:1991hy} that the Fradkin-Tseytlin term cancels the Weyl anomaly under the assumption of a constraint on the fermionic fields, which was argued to be necessary, because an analogous constraint was used to gauge-fix the fermions in the semi-light-cone gauge calculation of the one-loop effective action. 

In any case, the problem is that, without the dilaton, the Fradkin-Tseytlin counterterm is not available for general solutions of the GSE. This problem has been addressed recently. In \cite{Sakamoto:2017wor}, a counterterm has been proposed based on the doubled formalism for the type II case, which involves the Killing vector in combination with a dual coordinate. Another proposal was made in  \cite{Fernandez-Melgarejo:2018wpg} for the bosonic string. (It should also work for purely bosonic GSE solutions.) 

The purpose of this paper is to construct a generalized Fradkin-Tseytlin counterterm, which renders the GS superstring Weyl invariant. For simplicity, only the type I case will be considered. We will use old results \cite{Grisaru:1988wm, Grisaru:1988jt,Bellucci:1991hy} for the divergent one-loop effective action of the GS string to obtain the beta function and the Weyl anomaly, if no counterterm is included. The inspiration for the form of the counterterm comes from the calculation in \cite{Fernandez-Melgarejo:2018wpg} for the bosonic string as well as the geometry of superspace, which treats curvature and torsion on equal footing. Indeed, the counterterm we find involves the worldsheet torsion for a connection with vanishing curvature. Therefore, it cannot be expressed in terms of the metric and its derivatives, but it is nevertheless a covariant expression, both under diffeomorphisms and local Lorentz transformations. 
Another issue we address is the geometric interpretation of the vector $X_a$. A natural candidate for it is the torsion vector, $T_a=T_{ba}{}^b$, but this interpretation is not evident in the solution of \cite{Wulff:2016tju}, in which the bosonic torsion was set to zero. The interpretion of the three-form $H_{abc}$ as a torsion goes back to the classic work by Scherk and Schwarz \cite{Scherk:1974mc}, and similar ideas have been put forward for the dilaton \cite{Saa:1993mi}.\footnote{The dilaton has also been associated with non-metricity \cite{Dereli:1995zj,Vasilic:2006hs,Popovic:2007gm}, but one can expect, as metric affine gravity \cite{Hehl:1994ue, Hehl:2007bn} suggests, that non-metricity can be traded with torsion.} Therefore, we revisit the calculation by Tseytlin and Wulff for the type I case generalizing their solution to allow for arbitrary (bosonic) torsion and suggesting that one may uplift $X_a$ and $\chi_\alpha$ to the torsion supervector. Our solution is also useful in another respect. A common convention in the supergravity literature, and also in the papers on the superstring one-loop effective action, is a torsion constraint, in which the bosonic torsion is determined by the three-form. Our more general form of the GSE allows to translate between the different torsion constraints and makes the old results readily accessible. 

The rest of the paper is structured as follows. In Sec.~\ref{GSE}, we revisit the calculation by Tseytlin and Wulff and present the GSE in three distinct forms, the general case, the case with vanishing torsion, and the case of the standard supergravity torsion constraint. In Sec.~\ref{CT}, we consider the Weyl anomaly arising from the divergent one-loop effective action in the supergravity sector. Based on the GSE, we will construct a local expression in terms of $X_a$ and $\chi_\alpha$, which is equivalent to the Weyl anomaly modulo the classical field equations of the GS string. Then, we will write down the generalized Fradkin-Tseytlin counterterm. Sec.~\ref{conc} contains the conclusions, and the conventions for the gamma matrices are included in an appendix.

\section{Generalized supergravity equations from kappa-symmetry}
\label{GSE}
\subsection{Superspace Bianchi identities and kappa-symmetry}

In this section, we shall obtain the generalized supergravity equations. We closely follow the calculation by Tseytlin and Wulff \cite{Wulff:2016tju} and adopt also their notation. We will slightly deviate from them at the dimension 1 Bianchi identities by not constraining the bosonic torsion components to vanish.

In superspace, the torsion and curvature two-forms are defined by
\begin{equation}
\label{Bia:TR}
	T^A \equiv \nabla E^A = dE^A + E^B \wedge \Omega_B{}^A~,
	\qquad R_B{}^A = d\Omega_B{}^A +\Omega_B{}^C \wedge \Omega_C{}^A~.
\end{equation}
They satisfy the Bianchi identities
\begin{equation}
\label{Bia:BIs}	
	\nabla T^A = E^B \wedge  R_B{}^A~, \qquad 
	\nabla R_B{}^A =0~,
\end{equation}
or, in components,
\begin{align}
\label{Bia:TBI}
	\nabla_{[A} T_{BC]}{}^D + T_{[AB}{}^E T_{|E|C]}{}^D &= R_{[ABC]}{}^D~,\\
\label{Bia:RBI}
	\nabla_{[A} R_{BC]D}{}^E + T_{[AB}{}^F R_{|F|C]D}{}^E &= 0~.
\end{align}
The brackets denote graded commutation and include the normalization factor. 
We shall refer to \eqref{Bia:TBI} and \eqref{Bia:RBI} as the torsion Bianchi identity (TBI) and curvature Bianchi identity (RBI), respectively. 

It is a classical result \cite{Dragon:1978nf} that all curvature components are determined by the TBI in terms of the torsion and its covariant derivatives, because the curvature is a structure-group valued two-form, 
\begin{equation}
\label{Bia:Rstruct}
	R_{\alpha}{}^a = 0 = R_a{}^\alpha~,\qquad 
	R_\alpha{}^\beta = -\frac14 R_{ab} (\gamma^{ab})^\beta{}_\alpha = \frac14 R_{ab} (\gamma^{ab})_\alpha{}^\beta~.
\end{equation}
The RBI is then implied by virtue of the supergravity closure relations. 

The Bianchi identity for the three-form $H$ (HBI) reads, in components,
\begin{equation}
\label{Bia:HBI}
	\nabla_{[A} H_{BCD]} + \frac32 T_{[AB}{}^E H_{|E|CD]} = 0~.
\end{equation}

The strategy of Tseytlin and Wulff, which we follow, is to consider the superstring in the GS formalism as an embedding of the string worldsheet in superspace (for a review on superembeddings, see \cite{Sorokin:1999jx}). To remove spurious fermionic degrees of freedom, the superstring action must be invariant under kappa-symmetry transformations \cite{Shapiro:1986yy}. This constrains the background fields of dimension $-\frac12$ and $0$ to be \cite{Wulff:2016tju}
\begin{equation}
\label{I:Hminus12}
	H_{\alpha\beta\gamma}  =0
\end{equation}
and
\begin{equation}
\label{I:HT0}
	H_{a\alpha\beta}  = - i (\gamma_a)_{\alpha\beta}~, \qquad 
	T_{\alpha\beta}{}^a = - i (\gamma^a)_{\alpha\beta}~.
\end{equation}
In order to obtain the remaining components, one must solve the superspace Bianchi identities, which we will do next.

\subsection{Solution of the Bianchi identities}
The dimension zero HBI is implied by the Fierz identity \eqref{Bia:Fierz}.

\textbf{Dimension $\frac12$}.  The HBI and TBI, respectively, give rise to
\begin{align}
\label{I:HBI12}
	(\gamma^b)_{(\alpha\beta} \left( T_{\gamma)ab} + H_{\gamma)ab} \right) + (\gamma_a)_{\delta(\alpha} T_{\beta\gamma)}{}^\delta 
	&=0~,\\
\label{I:TBI12}	
	(\gamma^b)_{(\alpha\beta} T_{\gamma)b}{}^a - (\gamma^a)_{\delta(\alpha} T_{\beta\gamma)}{}^\delta 
	&=0~.
\end{align}
Adding these two equations yields 
\begin{equation}
\label{I:BI12.comb}
	(\gamma^b)_{(\alpha\beta} \left( H_{\gamma)ab} + 2 T_{\gamma)(ab)} \right) =0~. 
\end{equation}
One may adapt the frames and spin connections such that \cite{Wulff:2016tju}
\begin{equation}
\label{I:choice1}
	T_{\alpha[bc]} =0~, \qquad  (\gamma^b)^{\alpha\beta} T_{\beta ba} =0~.
\end{equation}
Under these conditions, the dimension $\frac12$ equations are solved by 
\begin{equation}
\label{I:HT12}
	H_{\alpha ab}  = 0~, \qquad 
	T_{\alpha a}{}^b=0~, \qquad
	T_{\alpha\beta}{}^\gamma = 2 \delta^\gamma_{(\alpha} \chi_{\beta)} - (\gamma^a)_{\alpha\beta} (\gamma_a \chi)^\gamma~,
\end{equation}
where $\chi_\alpha$ is an arbitrary (anti-commuting) spinor superfield. We note that it is the trace of the fermionic torsion,
\begin{equation}
\label{I:Tf.trace}
	T_{A \alpha}{}^A = T_{\beta \alpha}{}^\beta= 7 \chi_\alpha~.
\end{equation}

\textbf{Dimension $1$}. Here, we depart from \cite{Wulff:2016tju}.
The HBI reads
\begin{equation}
\label{I:HBI1}
	(\gamma^c)_{\alpha\beta} \left( T_{abc} + H_{abc} \right) - 2 (\gamma_{b})_{\gamma(\alpha} T_{\beta)a}{}^{\gamma}
	+2 (\gamma_{a})_{\gamma(\alpha} T_{\beta) b}{}^{\gamma} =0~, 
\end{equation}
while the TBI gives the two equations
\begin{align}
\label{I:TBI1.1}
	R_{\alpha\beta ab} &= -i (\gamma^c)_{\alpha\beta} T_{cab} -2i (\gamma_b)_{\gamma(\alpha} T_{|a|\beta)}{}^\gamma~,
\\
\label{I:TBI1.2}
	R_{(\alpha\beta\gamma)}{}^\delta &= \nabla_{(\alpha} T_{\beta\gamma)}{}^\delta 
	+ T_{(\alpha\beta}{}^\epsilon T_{|\epsilon|\gamma)}{}^\delta + T_{(\alpha\beta}{}^e T_{|e|\gamma)}{}^\delta~.
\end{align}
For the calculations, it is important to remember that the two-form $R_{ab}$ is antisymmetric and that the left hand sides of \eqref{I:TBI1.1} and \eqref{I:TBI1.2} are related by the $SO(1,9)$ structure relation \eqref{Bia:Rstruct}. 
We start by expanding $T_{a\alpha}{}^\beta$ into a basis of gamma-matrices,
\begin{equation}
\label{I:Tans}
	T_{a\alpha}{}^\beta = Y_a \delta_\alpha^\beta + \frac14 Z_{abc} (\gamma^{bc})_\alpha{}^\beta 
	+ Z_{abcde} (\gamma^{bcde})_\alpha{}^\beta~,
\end{equation}
with $Z_{abc}=Z_{a[bc]}$ and $Z_{abcde}=Z_{a[bcde]}$. Substituting \eqref{I:Tans} into \eqref{I:HBI1} and projecting the resulting expression onto the basis matrices symmetric in $\alpha\beta$ yields, from the $(\gamma^a)_{\alpha\beta}$ component,
\begin{equation}
\label{I:Tabc.sol}
	T_{abc} = - H_{abc} +2Z_{[ab]c} -4 Y_{[a} \eta_{b]c}~.
\end{equation}
The $(\gamma^{abcde})_{\alpha\beta}$ component yields, after some work,
\begin{equation}
\label{I:Zabcde}
	Z_{abcde}=0~.
\end{equation} 
Inserting these results into \eqref{I:TBI1.1}, one finds a term containing $Y_a$, which is not antisymmetric in $ab$. Therefore, we must conclude that
\begin{equation}
\label{I:Za}
	Y_{a}=0~.
\end{equation} 
In summary, \eqref{I:HBI1} and \eqref{I:TBI1.1} are solved by
\begin{align}
\label{I:Tbff}
	T_{a\alpha}{}^\beta &= \frac14 Z_{abc} (\gamma^{bc})_\alpha{}^\beta~,\\
\label{I:Tabc}
	T_{abc} &= - H_{abc} +2Z_{[ab]c}~,\\
\label{I:Rffbb}
	R_{\alpha\beta ab} &= i (\gamma^c)_{\alpha\beta} \left( H_{cab} - Z_{cab} \right)~.
\end{align}
Furthermore, one can use the index symmetries in \eqref{I:Tabc} to show that 
\begin{equation}
\label{I:Zabc}
	Z_{abc} = \frac12 H_{abc} + K_{abc}~,
\end{equation}
where
\begin{equation}
\label{I:contortion}
	K_{abc} = K_{a[bc]}= \frac12 \left( T_{abc} - T_{bca} + T_{cab} \right)
\end{equation}
is the contortion tensor. Hence, the arbitrariness of $Z_{abc}$ simply reflects the freedom to choose the bosonic connection. 

We now turn to \eqref{I:TBI1.2}. After substituting the previous results one finds that all the terms with $Z_{abc}$ cancel, so that the solution remains that of \cite{Wulff:2016tju},
\begin{equation}
\label{I:nab.f.chi}
	\nabla_{\alpha} \chi_\beta = \chi_\alpha \chi_\beta - \frac{i}{24}(\gamma^{abc})_{\alpha\beta} H_{abc}
	+ \frac{i}2 (\gamma^a)_{\alpha\beta} X_a~.
\end{equation}
The vector $X_a$ is arbitrary.

\textbf{Dimension $\frac32$.} The dimension-$\frac32$ component of the HBI is
\begin{equation}
\label{I:HBI32}
	\nabla_\alpha H_{abc} = 3 i (\gamma_{[a} \psi_{bc]})_\alpha~,
\end{equation}
where $\psi_{ab}^\alpha = T_{ab}{}^\alpha$ is the gravitino field strength.
The two TBIs of dimension $\frac32$ are 
\begin{align}
\label{I:TBI32.1}
	2 R_{\alpha[ab]c} &= \nabla_\alpha T_{abc} - i (\gamma_c \psi_{ab})_\alpha~,\\
\label{I:TBI32.2}
	2R_{a(\alpha\beta)}{}^\gamma &= \nabla_a T_{\alpha\beta}{}^\gamma + 2 \nabla_{(\alpha} T_{\beta) a}{}^\gamma
	+ T_{\alpha\beta}{}^b T_{ba}{}^\gamma + T_{\alpha\beta}{}^\delta T_{\delta a}{}^\gamma 
	+2 T_{a (\alpha}{}^\delta T_{\beta)\delta}{}^\gamma~. 
\end{align}
Using the identity 
\begin{equation}
\label{I:Rfbbb.ident}
	R_{\alpha bcd} = R_{\alpha[bc]d} + R_{\alpha[db]c} - R_{\alpha[cd]b}
\end{equation}
and the previous results, one obtains from \eqref{I:TBI32.1}
\begin{equation}
\label{I:Rfbbb}
	R_{\alpha abc} = \nabla_\alpha Z_{abc} - 2i (\gamma_{[b} \psi_{c]a})_\alpha~.
\end{equation}
From \eqref{I:TBI32.2}, after using \eqref{Bia:Rstruct} and \eqref{I:Rfbbb}, one finds after some work 
\begin{equation}
\label{I:nab.b.chi}
	\nabla_{a} \chi_\alpha = - \frac14 Z_{abc} (\gamma^{bc} \chi)_\alpha + \frac{i}2 (\gamma^b \psi_{ab})_\alpha~.
\end{equation}

\textbf{Dimension $2$.} The dimension-$2$ component of the HBI yields
\begin{equation}
\label{I:HBI2.sol}
	\nabla_{[a} H_{bcd]} = \frac32 H_{[ab}{}^e H_{cd]e} -3 Z_{[ab}{}^e H_{cd]e}~,
\end{equation} 
whereas the TBI contains two components of dimension 2,
\begin{align}
\label{I:TBI2.1}
	R_{[abc]d} &= \nabla_{[a} T_{bc]d} + T_{[ab}{}^e T_{|e|c]d}~,\\
\label{I:TBI2.2}
	R_{ab\alpha}{}^\beta &= \nabla_\alpha T_{ab}{}^\beta + 2 \nabla_{[a} T_{b]\alpha}{}^\beta 
	+T_{ab}{}^c T_{c\alpha}{}^\beta + T_{ab}{}^\gamma T_{\gamma\alpha}{}^\beta +2 T_{\alpha[a}{}^\gamma T_{|\gamma|b]}{}^\beta~.
\end{align}
Eq.~\eqref{I:TBI2.1} is just the usual bosonic torsion Bianchi identity. It becomes straightforwardly
\begin{align}
\label{I:Rabc.d}
	R_{[abc]d} &= -\nabla_{[a} H_{bc]d} +2\nabla_{[a} Z_{bc]d} + H_{[ab}{}^e H_{c]de} -2 Z_{[ab}{}^e H_{c]de} \\
\notag
	&\quad - H_{[ab}{}^e Z_{|e|c]d} - H_{[ab}{}^e Z_{c]de} + 2 Z_{[ab}{}^e Z_{|e|c]d} +2 Z_{[ab}{}^e Z_{c]de}~.
\end{align}
Obviously, $R_{a[bcd]}$ is determined by the identity
\begin{equation}
\label{I:Ra.bcd}
	R_{a[bcd]} = 2 R_{[abcd]} + R_{[bcd]a}~.
\end{equation}
Eq.~\eqref{I:TBI2.2} yields the spinor derivative of the gravitino field strength,\footnote{An alternative interpretation is that \eqref{I:TBI2.2} determines the curvature components $R_{abcd}$ \cite{Dragon:1978nf}.} 
\begin{align}
\label{I:nab.f.psi}
	\nabla_\alpha \psi_{ab}^\beta &= \chi_\alpha \psi_{ab}^\beta + \delta_\alpha^\beta (\chi \psi_{ab}) 
	- (\chi \gamma^c)^\beta (\gamma_c \psi_{ab})_\alpha\\
\notag &\quad
	+ \frac14 (\gamma^{cd})^\beta{}_\alpha \left(2 \nabla_{[a} Z_{b]cd} - R_{abcd} - H_{ab}{}^e Z_{ecd} 
	+ 2 Z_{[ab]}{}^e Z_{ecd} - 2 Z_{ace} Z_{bd}{}^e \right)~.
\end{align}

\textbf{Dimension $5/2$.} 
The remaining TBI yields

\begin{equation}
\label{I:TBI52}
	\nabla_{[a} \psi_{bc]}^\alpha  = - H_{[ab}{}^d \psi_{c]d}^\alpha + 2 Z_{[ab}{}^d \psi_{c]d}^\alpha 
	-\frac14 (\gamma^{de} \psi_{[ab})^\alpha Z_{c]de}~.
\end{equation}

\subsection{Closure of supersymmetry}

Having solved the TBI and HBI, we need to impose the closure of supersymmetry. This leads us to the generalized supergravity field equations.
 
Let us start with the Ricci identity
\begin{equation}
\label{I:nab.ff.chi}
	2\nabla_{(\alpha} \nabla_{\beta)} \chi_\gamma + T_{\alpha \beta}{}^\delta \nabla_\delta \chi_\gamma 
	+ T_{\alpha \beta}{}^a \nabla_a \chi_\gamma + R_{\alpha\beta \gamma}{}^\delta \chi_\delta =0~.
\end{equation}
After inserting the solutions of the Bianchi identities, using the Fierz identity and a bit of patience, one finds
\begin{equation}
\label{I:nab.f.X}
	\nabla_\alpha X_a = (\gamma_a \gamma^b \chi)_\alpha X_b + \frac1{12} (\gamma_a \gamma^{bcd} \chi)_\alpha H_{bcd} 
	-\frac{i}4 (\gamma_a{}^{bc} \psi_{bc})_\alpha~.
\end{equation}
Notice again that terms with $Z_{abc}$ have cancelled, and the result is identical to that of \cite{Wulff:2016tju}.

The next identity we consider is 
\begin{equation}
\label{I:nab.fb.chi}
	2\nabla_{[a} \nabla_{\alpha]} \chi_\beta + T_{a\alpha}{}^\gamma \nabla_\gamma \chi_\beta 
	+ T_{a \alpha}{}^b \nabla_b \chi_\beta + R_{a\alpha\beta}{}^\gamma \chi_\gamma =0~.
\end{equation}
After a bit of algebra, this becomes 
\begin{align}
\notag 
	&(\gamma^b)_{\alpha\beta} \left( 2 \nabla_a X_b - 2 \chi \psi_{ab} + 2 Z_{abc} X^c + R_{ab} + \nabla_a Z_b 
	- \nabla_c Z_{ab}{}^c - H_{ac}{}^d Z_{db}{}^c - Z_{ca}{}^d Z_{db}{}^c - Z_{abc} Z^c \right)\\
\notag
	&-\frac12 (\gamma^{bcd})_{\alpha\beta} \left( \frac13 \nabla_a H_{bcd} - \nabla_a Z_{bcd} + \nabla_b Z_{acd} 
	+ R_{abcd} \right. \\
\label{I:bf.chi.closure}
	&\left. \phantom{\frac12} + Z_{abe} H_{cd}{}^e + H_{ab}{}^e Z_{ecd} - Z_{ab}{}^e Z_{ecd} + Z_{ba}{}^e Z_{ecd} 
	+ 2 Z_{ace} Z_{bd}{}^e 	\right) =0~.
\end{align}
Here, we have defined 
\begin{equation}
\label{I:Z.a}
	Z_a = Z_{ba}{}^b~,
\end{equation}
which equals the torsion vector, because of \eqref{I:Zabc} and \eqref{I:contortion},
\begin{equation}
\label{I:Tb.trace}
	T_a = T_{ba}{}^b = Z_a~.
\end{equation}

The antisymmetric part in \eqref{I:bf.chi.closure}, which comprises the terms on the second and third lines, vanishes identically by \eqref{I:Ra.bcd} and \eqref{I:Rabc.d}. The remaining stuff yields the field equation
\begin{equation}
\label{I:nab.b.X}
	R_{ab} - 2 \chi \psi_{ab} + 2 \nabla_a X_b + 2 Z_{abc} X^c +  \nabla_a Z_b 
	- \nabla_c Z_{ab}{}^c - H_{ac}{}^d Z_{db}{}^c - Z_{ca}{}^d Z_{db}{}^c + Z_{abc} Z^c =0~.
\end{equation}
Combining its antisymmetric part with \eqref{I:Rabc.d} and using the identity 
\begin{equation}
\label{I:Ricci.asym}
	R_{[ab]} = R_{c[ab]}{}^c = \frac32 R_{[cab]}{}^c
\end{equation}
leads to
\begin{equation}
\label{I:nab.b.X.asym}
	\nabla_{[a} X_{b]} + Z_{[ab]c} X^c  +\frac12 Z_{d[a}{}^c H_{b]c}{}^d 
	- \frac14 \nabla_c H_{ab}{}^c + \frac14 H_{abc} Z^c = \chi \psi_{ab}~.
\end{equation}

Finally, consider the Ricci identity
\begin{equation}
\label{I:nab.ff.X}
	2\nabla_{(\alpha} \nabla_{\beta)} X_a + T_{\alpha \beta}{}^\delta \nabla_\delta X_a 
	+ T_{\alpha \beta}{}^b \nabla_b X_a + R_{\alpha\beta a}{}^b X_b =0~.
\end{equation}
Using the previous results and quite a bit more of patience, one finds
\begin{equation}
\label{I:nab.b.X.trace}
	\nabla_a X^a - 2 X_a X^a - Z_a X^a + \frac1{12} H_{abc} H^{abc} = \frac{i}3 H_{abc} (\chi \gamma^{abc} \chi) +
	(\chi \gamma^{ab} \psi_{ab})~.
\end{equation}
This completes the closure relations.

\subsection{Other forms of the field equations}
\label{OF}

Compared to \cite{Wulff:2016tju}, our field equations allow for an arbitrary bosonic torsion. It is contained in the tensor $Z_{abc}$, which has the same index structure as the contortion tensor. This means that our equations represent the GSE for an arbitrary choice of bosonic connections. Vice versa, this freedom can be used to relate our equations to various choices of torsion constraints that have been used in the literature. 

We recall that the contortion tensor \eqref{I:contortion} can be used to express the bosonic connection and curvature in terms of the unique bosonic (torsion-free) connection ($\bar{\nabla}_a$) and the Riemann curvature tensor ($\bar{R}_{abcd}$). This is achieved by the relations
\begin{align}
\label{I:nab.torsion.free}
	\nabla_a X_b &= \bar{\nabla}_a X_b - K_{ab}{}^c X_c~, \\
\label{I:nab.torsion.free.2}
	\nabla_a \chi_\alpha &= \bar{\nabla}_a \chi_\alpha - \frac14 K_{abc} (\gamma^{bc}\chi)_\alpha~,\\
\label{I:R.torsion.free}
	R_{abcd} &= \bar{R}_{abcd} + 2 \bar{\nabla}_{[a} K_{b]cd} + 2 K_{[a|c|}{}^e K_{b]de}~.
\end{align}
Eliminating also $Z_{abc}$ by \eqref{I:Zabc}, the equations \eqref{I:HBI2.sol} and \eqref{I:Rabc.d} reduce to the usual bosonic Bianchi identities
\begin{equation}
\label{I:Bianchi.torsion.free}
	\bar{\nabla}_{[a} H_{bcd]} =0~,\qquad \bar{R}_{[abc]d} =0~,
\end{equation}
respectively, while \eqref{I:nab.b.X}, \eqref{I:nab.b.X.asym} and \eqref{I:nab.b.X.trace} become
\begin{align}
\label{I:Ricci.torsion.free}
	\bar{R}_{ab} + 2 \bar{\nabla}_{(a} X_{b)} - \frac14 H_{a}{}^{cd} H_{bcd} &=0~,\\
\label{I:nab.b.X.asym.torsion.free}	
	\bar{\nabla}_{[a} X_{b]} -\frac14 \bar{\nabla}^c H_{abc} + \frac12 H_{abc} X^c &= \chi \psi_{ab}~,\\
\label{I:nab.b.X.trace.torsion.free}  
	\bar{\nabla}_{a} X^a -2 X_a X^a +\frac1{12} H^{abc} H_{abc} &= \frac{i}3 H_{abc}\, \chi \gamma^{abc} \chi + 
	\chi \gamma^{ab} \psi_{ab}~. 
\end{align}
Moreover, \eqref{I:nab.b.chi}, \eqref{I:nab.f.psi} and \eqref{I:TBI52} give rise to 
\begin{align}
\label{I:nab.b.chi.torsion.free}
	\bar{\nabla}_a \chi_\alpha &=-\frac18 H_{abc} (\gamma^{bc} \chi)_\alpha - \frac{i}2 (\gamma^b \psi_{ab})_\alpha~,\\
\label{I:nab.f.psi.torsion.free}
	\nabla_\alpha \psi_{ab}^\beta &= \chi_\alpha \psi_{ab}^\beta + \delta_\alpha^\beta (\chi \psi_{ab}) 
	- (\chi \gamma^c)^\beta (\gamma_c \psi_{ab})_\alpha
	+ \frac14 (\gamma^{cd})^\beta{}_\alpha \left( \bar{\nabla}_{[a} H_{b]cd} - \bar{R}_{abcd} - \frac12 H_{ac}{}^e H_{bde}	
	\right)~,\\
\label{I:TBI52.torsion.free}
	\bar{\nabla}_{[a} \psi_{bc]}^\alpha  &= -\frac18 (\gamma^{de} \psi_{[ab})^\alpha H_{c]de}~.
\end{align}
Eqs.~\eqref{I:Bianchi.torsion.free}--\eqref{I:TBI52.torsion.free} are, of course, just the field equations obtained in \cite{Wulff:2016tju}. The contortion tensor has dropped out everywhere, which could have been anticipated from the fact that setting $K_{abc}=0$ is a gauge choice. 

In most of the older literature \cite{Witten:1985nt,Atick:1985de}, in particular in the papers on the quantization of the GS superstring \cite{Grisaru:1988jt,Grisaru:1988wm,Grisaru:1988sa,Majumdar:1989sx,Bellucci:1991hy}, which we wish to use to discuss the Weyl anomaly cancellation, a torsion constraint is adopted that corresponds to the gauge choice $Z_{abc}=0$ in our notation. We will use the symbol `` $\mathring{}$ '' to distinuish this choice from the general case. As is evident from \eqref{I:Tabc}, the bosonic connection $\mathring{\nabla}_a$ has a totally antisymmetric torsion,
\begin{equation}
\label{OF:Tabc}
	\mathring{T}_{abc} = - H_{abc}~.
\end{equation}
The generalized supergravity equations take the form 
\begin{align}
\label{OF:HBI2.sol}
	\mathring{\nabla}_{[a} H_{bcd]} &= \frac32 H_{[ab}{}^e H_{cd]e}~,\\
\label{OF:Rabc.d}
	\mathring{R}_{[abc]d} &= -\mathring{\nabla}_{[a} H_{bc]d} + H_{[ab}{}^e H_{c]de}~, \qquad
	\mathring{R}_{[ab]} = -\frac12 \mathring{\nabla}_c H_{ab}{}^c~, \\
\label{OF:R.ab}
	\mathring{R}_{ab} + 2 \mathring{\nabla}_a X_b &= 2 \chi \psi_{ab}~,\\
\label{OF:nab.b.X.trace}
	\mathring{\nabla}_a X^a - 2 X_a X^a + \frac1{12} H_{abc} H^{abc} &= \frac{i}3 H_{abc} (\chi \gamma^{abc} \chi) +
	(\chi \gamma^{ab} \psi_{ab})~,\\
\label{OF:nab.b.chi}
	\mathring{\nabla}_{a} \chi_\alpha &= \frac{i}2 (\gamma^b \psi_{ab})_\alpha~,\\
\label{OF:nab.f.psi}
	\mathring{\nabla}_\alpha \psi_{ab}^\beta &= \chi_\alpha \psi_{ab}^\beta + \delta_\alpha^\beta (\chi \psi_{ab}) 
	- (\chi \gamma^c)^\beta (\gamma_c \psi_{ab})_\alpha - \frac14 \mathring{R}_{abcd} (\gamma^{cd})^\beta{}_\alpha~,\\
\label{OF:TBI52}
	\mathring{\nabla}_{[a} \psi_{bc]}^\alpha  &= - H_{[ab}{}^d \psi_{c]d}^\alpha~.
\end{align}

Yet a diffent torsion constraint was adopted in \cite{Bonora:1986xd,Candiello:1993di} by imposing $R_{\alpha\beta}=R_{\gamma\alpha\beta}{}^\gamma=0$. In our notation, this would correspond to $Z_{abc}=T_{abc}=H_{abc}$. We will not give the details for this choice, as we will not need them. 

The interpretation of the three-form as a torsion goes back to the classic work by Scherk and Schwarz \cite{Scherk:1974mc}. It is, however, evident that the torsion tensor $T_{abc}$ has enough degrees of freedom to accomodate not only $H_{abc}$, but also $X_a$. Indeed, the torsion vector $T_a=Z_a$ is really its natural place. Therefore, in our opinion, a torsion constraint that relates the antisymmetric part of the torsion to $H_{abc}$ and the torsion vector to $X_a$ is preferrable, because it gives them a precise geometric meaning. In the supergravity case, such a choice was advocated, e.g., in \cite{Saa:1993mi}. For example, one could use the constraint
\begin{equation}
\label{I:gauge.choice}
	R_{\gamma\alpha\beta}{}^\gamma -\frac12 \nabla_A T_{\alpha \beta}{}^A =0~.
\end{equation} 
This would imply 
\begin{equation}
\label{I:gauge.choice.expl}
	Z_a = 7 X_a~, \qquad Z_{[abc]} =H_{abc}~,
\end{equation}
while one can set all the remaining components of $T_{abc}$ to zero. Then, with \eqref{I:Tf.trace} and \eqref{I:Tb.trace}, the supertorsion vector is simply $T_A = 7 ( X_a, \chi_\alpha )$.\footnote{It is tempting to try to uplift some of the field equations into superspace as a ``superspace Bianchi identity'' for the torsion supervector, as was done in Sec.~4 of \cite{Wulff:2016tju} for the type-IIB case. Although this works for the $\alpha\beta$ and $\alpha a$ components, we were not able to incorporate the $ab$ components, because the type I equations are different from type II case.}

\section{Generalized Fradkin-Tseytlin counter term}
\label{CT}

In this section, we will construct the generalized Fradkin-Tseytlin counter term, which renders the superstring sigma model 
Weyl invariant at the one-loop level in the supergravity sector. We recall that the one-loop terms from the gauge sector of the heterotic string are of the same degree in $\alpha'$ as two-loop supergravity terms \cite{Grisaru:1988sa}.

We start with the classical action in superspace
\begin{equation}
\label{CT:action}
	S = -\frac1{4\pi \alpha'} \int \rmd^2 \xi \sqrt{-G} \left[ G^{IJ} \eta_{ab}(\partial_I z^M) E_M{}^a (\partial_J z^N) E_N{}^b 
	-\epsilon^{IJ} (\partial_I z^M) (\partial_J z^N) B_{NM} \right]~, 
\end{equation}
where $\xi^I$ ($I, J=0,1$) are the worldsheet coordinates. We treat $G_{IJ}$ as an independent worldsheet metric that will be fixed later, by its field equation and exploiting the Weyl symmetry of the action \eqref{CT:action}, to the induced metric
\begin{equation}
\label{CT:metric}
	G_{IJ} = E_I{}^a E_J{}^b \eta_{ab}~, \qquad E_I{}^A = (\partial_I z^M) E_M{}^A~,
\end{equation}
where $z^M=(x^m, \theta^\mu)$ are the superspace coordinates. Morever, $\epsilon^{IJ}$ denotes the covariant epsilon tensor.

Assuming that the background satisfies the generalized supergravity equations derived in the previous section, the variation of the action \eqref{CT:action} under variations of $z^M(\xi)$ is found as 
\begin{align}
\label{CT:var.action}
	\delta S &= \frac1{2\pi \alpha'} \int \rmd^2 \xi \sqrt{-G} \left\{ -i \delta z^M E_M{}^\alpha
		 (1-\Gamma)_{\alpha}{}^{\beta} G^{IJ} E_J{}^a (\gamma_a)_{\beta \gamma} E_I{}^\gamma \phantom{\frac12} \right. \\
\notag 	&\quad \left.	
	+ \delta z^M E_M{}^a 
		\left[ G^{IJ} \left( D_I E_{Ja} - E_I{}^c E_J{}^b T_{a(bc)} \right)
		-\frac12 \epsilon^{IJ} \left( H_{abc} E_I{}^b E_J{}^c 
		+ i (\gamma_a)_{\alpha\beta} E_I{}^\alpha E_J{}^\beta \right)\right]  \right\}~,
\end{align}
where 
\begin{equation}
\label{CT:D.def}
	D_I E_{Ja} = \partial_I E_{Ja} - \bar{\Gamma}^K{}_{IJ} E_{Ka} + (\partial_I z^M) \Omega_{Ma}{}^b E_{Jb}~,
\end{equation}
with $\bar{\Gamma}^K{}_{IJ}$ being the Christoffel symbols (torsion-free connection) on the world sheet, and $\Omega_{Ma}{}^b$ the  superspace spin connections. Furthermore, $\Gamma$ denotes the matrix
\begin{equation}
\label{CT:gamma}
	\Gamma=\frac12 \epsilon^{IJ} E_I{}^a E_J{}^b \gamma_{ab}~.
\end{equation}
The action is evidently invariant under the $\kappa$-symmetry transformations
\begin{equation}
\label{CT:kappa}
	\delta z^M E_M{}^a =0~,\qquad  \delta z^M E_M{}^\alpha = \frac12 \kappa^\beta (1+\Gamma)_{\beta}{}^{\alpha}~. 
\end{equation}
The remaining field variations in \eqref{CT:var.action} yield the classical field equations
\begin{align}
\label{CT:fe.b}
	G^{IJ} \bar{D}_I E_{Ja} -\frac12 \epsilon^{IJ} \left( H_{abc} E_I{}^b E_J{}^c 
		+ i (\gamma_a)_{\alpha\beta} E_I{}^\alpha E_J{}^\beta \right) &=0~,\\
\label{CT:fe.f}	
	(1-\Gamma)_{\alpha}{}^{\beta} G^{IJ} E_J{}^a (\gamma_a)_{\beta \gamma} E_I{}^\gamma &=0~.
\end{align}
In \eqref{CT:fe.b}, we have absorbed the torsion term into the covariant derivative using
\begin{equation}
\label{CT:D.D.bar}
	D_I E_{Ja} - E_I{}^c E_J{}^b T_{a(bc)} = \bar{D}_I E_{Ja}~,
\end{equation}
where
\begin{equation}
\label{CT:D.bar.def} 
	\bar{D}_I E_{Ja} = \partial_I E_{Ja} - \bar{\Gamma}^K{}_{IJ} E_{Ka} + (\partial_I z^M) \bar{\Omega}_{Ma}{}^b E_{Jb}
\end{equation}
contains the (unique) torsion-free spin connection in the bosonic components of $\bar{\Omega}_{Ma}{}^b$.

The Weyl anomaly of the supersymmetric sigma model is proportional to the beta functions for the metric and $B$-field, which can be read off from the divergent terms of the one-loop effective action \cite{Grisaru:1988wm}. It is given by 
\begin{equation}
\label{CT:anomaly}
	\vev{T^I{}_I} = \frac12 \left( G^{IJ} + \epsilon^{IJ} \right) \left( E_I{}^a E_J{}^b \mathring{R}_{ab} + 
		E_I{}^a E_J{}^\alpha \mathring{R}_{\alpha a} \right)~,
\end{equation}
where we have retained the torsion constraint $Z_{abc}=0$ that was used in the original paper. 

In order to construct a suitable counterterm, our first aim is to find an expression, which is equivalent to the right hand side of \eqref{CT:anomaly} modulo the classical field equations \eqref{CT:fe.b} and \eqref{CT:fe.f}. Consider 
\begin{equation}
\label{CT.term1}
	\bar{\nabla}_I \left( G^{IJ} E_J{}^a X_a \right) = G^{IJ} (\bar{D}_I E_J{}^a) X_a 
		+ G^{IJ} E_J{}^a (E_I{}^b \bar{\nabla}_b X_a + E_I{}^\alpha \bar{\nabla}_\alpha X_a )~.  
\end{equation}
Using \eqref{CT:fe.b} and the GSE for the background fields, we get
\begin{align}
\label{CT:term1.1}
	\bar{\nabla}_I \left( G^{IJ} E_J{}^a X_a \right) &= G^{IJ} E_J{}^a E_I{}^b \left(-\frac12 \bar{R}_{ab} 
	+ \frac18 H_{a}{}^{cd} H_{bcd} \right) + G^{IJ} E_I{}^\alpha E_J{}^b \left(-\frac12 \bar{R}_{\alpha b} 
	+ (\gamma_a \Xi)_\alpha \right)\\
\notag
&\quad 
	+ \frac12 \epsilon^{IJ} \left( E_I{}^b E_J{}^c H_{abc} 
	+ i (\gamma_a)_{\alpha\beta} E_I{}^\alpha E_J{}^\beta \right) X^a~,
\end{align}  
where we have introduced
\begin{equation}
\label{CT:Xi.def}
	\Xi^\alpha = (\gamma^a \chi)^\alpha X_a +\frac1{12} (\gamma^{abc}\chi)^\alpha H_{abc} 
	- \frac{i}4 (\gamma^{ab}\psi_{ab})^\alpha~.
\end{equation}

Similarly, one has
\begin{align}
\notag
	\bar{\nabla}_I \left( \epsilon^{IJ} E_J{}^a X_a \right) &= \epsilon^{IJ} E_J{}^a \left( E_I{}^b \nabla_b X_a 
	+ E_I{}^\alpha \nabla_\alpha X_a +\frac12 T_{[ba]c} X^c \right)
	+\frac{i}2 \epsilon^{IJ} (\gamma_a)_{\alpha\beta} E_I{}^\alpha E_J{}^\beta  X^a\\
\label{CT:term2}
	&= \epsilon^{IJ} E_I{}^a E_J{}^b \left( \frac14 \bar{\nabla}_c H_{ab}{}^c -\frac12 H_{abc}X^c +\chi \psi_{ab} \right) \\
\notag &\quad
+\epsilon^{IJ} E_J{}^a E_I{}^\alpha \left(-\frac12 \bar{R}_{\alpha a} 
	+ (\gamma_a \Xi)_\alpha \right) + \frac{i}2 \epsilon^{IJ} (\gamma_a)_{\alpha\beta} E_I{}^\alpha E_J{}^\beta X^a~.
\end{align}
and
\begin{align}
\notag
	\bar{\nabla}_I \left( \epsilon^{IJ} E_J{}^\alpha \chi_\alpha \right) &= \epsilon^{IJ} \left( 
		E_I{}^a E_J{}^\alpha \nabla_a \chi_\alpha + E_J{}^\beta E_I{}^\alpha \nabla_\alpha \chi_\beta 
		+\frac12 E_J{}^A E_I{}^B T_{BA}{}^\alpha \chi_\alpha \right) \\
\label{CT:term3}
	&= -\frac12 \epsilon^{IJ} \left( -E_J{}^a E_I{}^\alpha \bar{R}_{\alpha a} +E_I{}^a E_J{}^b \chi \psi_{ab} 
	+i (\gamma_a)_{\alpha\beta} E_I{}^\alpha E_J{}^\beta X^a \right)~.
\end{align}
Now, one can combine \eqref{CT:term1.1}, \eqref{CT:term2} and \eqref{CT:term3} into
\begin{equation}
\label{CT:all.terms}
	\bar{\nabla}_I \left( G^{IJ} E_J{}^a X_a + \epsilon^{IJ} E_J{}^a X_a  +2 \epsilon^{IJ} E_J{}^\alpha \chi_\alpha \right) 
	= -\frac12 \left( G^{IJ} +\epsilon^{IJ}\right) \left( E_I{}^a E_J{}^b \mathring{R}_{ab} + 
		E_I{}^a E_J{}^\alpha \mathring{R}_{\alpha a} \right)~,
\end{equation}
where the terms containing $\Xi^\alpha$ have cancelled by virtue of the fermionic field equation \eqref{CT:fe.f}. Moreover, we have translated the curvatures to the torsion constraint $Z_{abc}=0$, which readily exposes the Weyl anomaly \eqref{CT:anomaly} on the right-hand side of \eqref{CT:all.terms}.

The generalized Fradkin-Tseytlin counterterm we are looking for must be such that its classical contribution to the trace of the worldsheet stress-energy tensor equals the left hand side of \eqref{CT:all.terms}, cancelling the one-loop Weyl 
anomaly. For this purpose, we need to discuss some aspects of torsion in two dimensions. In 2-$d$, the torsion tensor has only two indepenent components, which are the components of the torsion vector. Therefore, the contortion tensor \eqref{I:contortion} is of the general form 
\begin{equation}
\label{CT:contortion.2d}
	K_{IJK} = G_{IK} T_J - G_{IJ} T_K~.
\end{equation}
Furthermore, the general curvature is related to the Riemann curvature tensor by [cf.\ \eqref{I:R.torsion.free}] 
\begin{align}
\label{CT:R4}
	R_{IJKL} &= \bar{R}_{IJKL} + \left( \bar{\nabla}_I T_K -T_I T_K \right) G_{JL} 
	- \left( \bar{\nabla}_I T_L -T_I T_L \right) G_{JK} \\
\notag &\quad
	+ \left( \bar{\nabla}_J T_L -T_J T_L \right) G_{IK} 
	- \left( \bar{\nabla}_J T_K -T_J T_K\right) G_{IL} 
	+ \left( G_{IK} G_{JL} - G_{JK} G_{IL} \right) T_M T^M~.
\end{align}
Taking the trace of \eqref{CT:R4}, one finds 
\begin{equation}
\label{CT:R2}
	R_{IJ} = \bar{R}_{IJ} - G_{IJ} \bar{\nabla}_K T^K~, \qquad R = \bar{R} - 2 \bar{\nabla}_I T^I~.
\end{equation}
Therefore, if we adopt a curvature-free connection $\tilde{\Omega}_I{}^J$, then
\begin{equation}
\label{CT:tilde.R}
	\tilde{R}_{IJKL}=0 : \qquad \bar{R}_{IJ} = G_{IJ} \bar{\nabla}_K \tilde{T}^K~, 
	\qquad \bar{R} = 2 \bar{\nabla}_I \tilde{T}^I~.
\end{equation}
Furthermore, we know that
\begin{equation}
\label{CT:R.prop}
	\bar{R}_{IJKL} = \frac12 ( G_{IL} G_{JK} - G_{IL} G_{JK} ) \bar{R}~.
\end{equation}
Substituting \eqref{CT:R.prop} into \eqref{CT:R4} and setting the left-hand side to zero shows that $\tilde{T}_I$ must satisfy
\begin{equation}
\label{CT:T.prop} 
	\bar{\nabla}_{I} \tilde{T}_J - \tilde{T}_I \tilde{T}_J = \Lambda G_{IJ}
\end{equation}
for some $\Lambda$. Clearly, this implies
\begin{equation}
\label{CT:T.prop2}
	\bar{\nabla}_{[I} \tilde{T}_{J]} =0~.
\end{equation}

The simplest representative of a torsion-free connection is obtained, of course, for vanishing spin connections. In this case, one has 
\begin{equation}
\label{CT:T.expl}
	\tilde{T}_I = - \frac1{e} e_{I}{}^i \partial_J (e e_i{}^J)\qquad (\Omega_{Iij}=0)~,
\end{equation}
with the zweibein $e_I{}^i$, inverse zweibein $e_i{}^I$, and $e =\det(e_I{}^i)$. It is interesting to note that this expression coincides with the construction of the counterterm in \cite{Fernandez-Melgarejo:2018wpg}. Indeed, if we denote by $\bar{\Omega}_I{}^{ij}$ the unique torsion-free spin connection, then we easily verify that
\begin{equation}
\label{CT:verify}
	\bar{\Omega}_{i}{}^{ij} = e_i{}^I \bar{\Omega}_I{}^{ij} = \frac1{e} \partial_I( e e^{jI})~.
\end{equation}
Clearly, a definition in terms of the spin connection is not covariant under local Lorentz frame rotations. By the same token, defining $\tilde{T}_I$ by \eqref{CT:T.expl} would make it covariant only under global Lorentz frame rotations, not under local ones. 
However, this is not what we have in mind. We define $\tilde{T}_I$ as the torsion vector for an arbitrary curvature-free connection. Therefore, it transforms covariantly under diffeomorphisms and is actually invariant under local Lorentz frame rotations.\footnote{Remember that the spin connection changes under local Lorentz frame rotations. This does not affect the property of vanishing curvature.}

We can now write down the generalized Fradkin-Tseytlin counterterm. Defining 
\begin{equation}
\label{CT:S.c}
	S_\mathrm{c} = -\frac1{2\pi} \int \rmd^2 \xi \, e \tilde{T}_I \left( G^{IJ} E_J{}^A \Phi_A + \epsilon^{IJ} E_J{}^A \Psi_A \right)  
\end{equation}
with two supervectors $\Phi_A$ and $\Psi_A$, the worldsheet stress-energy tensor receives a contribution
\begin{equation}
\label{CT:T.c}
	\vev{T^I{}_I}_\mathrm{c} = \frac{2\pi}{e} e_i{}^{I} \frac{\delta S_\mathrm{c}}{\delta e_i{}^{I}} =
	\bar{\nabla}_I \left( G^{IJ} E_J{}^A \Phi_A + \epsilon^{IJ} E_J{}^A \Psi_A \right)~.
\end{equation}
Therefore, from \eqref{CT:all.terms} and \eqref{CT:anomaly} we see that for 
\begin{equation}
\label{CT:Phi.Psi}
	\Phi_A = (X_a, 0)~,\qquad \Psi_A = (X_a, 2\chi_\alpha)
\end{equation}
the counterterm cancels the one-loop Weyl anomaly.

The counterterm \eqref{CT:S.c} cannot be written as a local functional of the worldsheet metric and its derivatives. One can see this   as follows.\footnote{I thank A.~Tseytlin for this elegant derivation.} Eq.~\eqref{CT:T.prop2} implies that $\tilde{T}_I$ can be locally written as the gradient of some scalar, $\tilde{T}_I=\bar{\nabla}_I \omega$. This scalar, because of \eqref{CT:tilde.R}, must satisfy
\begin{equation}
\label{CT:T.nonloc}
	\bar{R} = 2 \bar{\nabla}^2 \omega~,
\end{equation}
so that $\tilde{T}_I$ is non-local in the metric. We will comment on this fact in the conclusions.

It is instructive to consider the supergravity case, for which $X_a = \nabla_a \Phi$ and $\chi_\alpha = \nabla_\alpha \Phi$, with $\Phi$ being the dilaton. In this case, the counter term \eqref{CT:T.c} can be written as
\begin{equation}
\notag
	S_\mathrm{c} = -\frac1{2\pi} \int \rmd^2 \xi\, e \tilde{T}_I \left[ (G^{IJ} + \epsilon^{IJ}) \partial_J \Phi 
	- (G^{IJ} - \epsilon^{IJ}) E_J{}^\alpha \nabla_\alpha \Phi \right]~.
\end{equation}
Integrating by parts the term with $\partial_J \Phi$ and using \eqref{CT:tilde.R} and \eqref{CT:T.prop2}, one finds
\begin{equation}
\notag
	S_\mathrm{c} = \frac1{4\pi} \int \rmd^2 \xi\, e \left[ \bar{R} \Phi 
	+ 2\tilde{T}_I  (G^{IJ} - \epsilon^{IJ}) E_J{}^\alpha \nabla_\alpha \Phi \right]~.
\end{equation}
The first term in the brackets is the Fradkin-Tseytlin counter term. The remaining term vanishes identically, if one imposes an additional constraint on the fermionic background \cite{Bellucci:1991hy}.\footnote{Cf.\ (5.6) of \cite{Bellucci:1991hy}. The apparent difference in the sign of the term with the epsilon tensor can be traced back to the same difference between their (4.2) and our \eqref{CT:action}.} This constraint was motivated with the argument that the one-loop effective action was calculated in a semi-light-cone gauge, in which the constraint represents the gauge fixing for the fermionic fluctuations. Accordingly, the same constraint should be used for the background. 
Our results show that this artifact disappears for the generalized Fradkin-Tseytlin counter term.

\section{Conclusions}
\label{conc}

In this paper, we have revisited the recent derivation of the GSE based upon the requirement of invariance of the GS sigma model under kappa-symmetry transformations. Compared to the solution given by Tseytlin and Wulff, we have allowed for an arbitrary bosonic torsion, which simply reflects the freedom of choice of the bosonic connections. Our more general solution is useful for a comparison with other torsion constraints in the supergravity literature and enables us to interpret the vector $X_a$ as a torsion vector, which naturally forms a torsion supervector together with the dilatino $\chi_\alpha$. Our main result is the construction of the generalized Fradkin-Tseytlin counterterm, which makes the GS string Weyl invariant at the one-loop level in the supergravity sector. Interestingly, the new counterterm does not feel the ambiguity of the additional constraint on the fermionic background fields. This ambiguity was shown to be an artifact of the standard Fradkin-Tseytlin term. 

Despite the formal cancellation of the Weyl anomaly, the counterterm has to be taken with a grain of salt. A hint that something is amiss comes from the fact that the counterterm cannot be written as a local functional of the worldsheet metric. In fact, the torsion vector introduces a new degree of freedom. In our treatment, which takes the zweibein and the spin connection as independent variables, this new field is the spin connection, which is necessary to retain covariance under local Lorentz frame rotations. The spin connection is taken to be invariant under Weyl transformations, otherwise the restriction to a curvature-free connection would not make sense.\footnote{In two dimensions, a curvature-free spin connection can be locally parameterized by a scalar.} The field equation of the spin connection, however, would impose an equation, which is not implied by the GSE and the classical string field equations. Moreover, the same field equation would render the Ward identity for local Lorentz frame rotations anomalous.
Therefore, one ends up in the strange situation of a field, for which one cannot impose its field equation. 
An alternative viewpoint on torsion is to take the metric and the contortion tensor as independent variables. In this approach, the torsion vector could be taken as invariant under Weyl transformations, but then our counterterm would not at all cancel the trace anomaly. However, as mentioned above, with such transformation properties one cannot impose a curvature-free connection, because the curvature would not be Weyl invariant. 
A formally simple way of obtaining a local counterterm is to introduce the scalar field $\omega$, set $\tilde{T}_I=\partial_I \omega$ in \eqref{CT:S.c} and impose the relation \eqref{CT:T.nonloc} by means of a Lagrange multiplier field. For consistency, $\omega$ transforms by a shift under Weyl transformations,\footnote{$G_{IJ}\to \e{2\alpha} G_{IJ}$ requires $\omega \to \omega -\alpha$.}
while the Lagrange multiplier is invariant. Hence, the worldsheet stress-energy tensor would not be traceless, but the Ward identity for Weyl transformations would be maintained by the transformation of $\omega$.
In conclusion, none of the above alternatives is really satisfactory, and it remains unclear whether a general GSE background can be considered on equal footing with supergravity backgrounds. We suspect that the problem is related to the fact the GSE are not truely field equations. (There are more fields than equations). We leave this interesting issue open for debate.

The debut of the torsion (super)vector raises the interesting possibility to reformulate (generalized) supergravity entirely in terms of curvature and torsion. Also, it is not clear whether or not the GSE may be obtained from an action principle. One should not expect that the GSE correspond to some kind of simple torsion gravity. It is well known that all gravitational actions containing terms with up to two derivatives (\ie linear in curvature, quadratic in torsion, or with a single derivative of torsion), without matter fields, give descriptions equivalent to Einstein gravity. A related question is the uplift to superspace. On the one hand, we have suggested that, with a suitable torsion constraint, $X_a$ and $\chi_\alpha$ combine into the torsion supervector $T_A$. On the other hand, the structure of the counterterm suggests that there are two relevant supervectors, \eg $\phi_A$ and $\psi_A$ of \eqref{CT:Phi.Psi}.

For simplicity, we have considered here only the type I case. We expect that the type II cases can be treated in a similar fashion. 
Moreover, it would be interesting to investigate how the GSE are affected by $\alpha'$ corrections, in analogy to the supergravity equations \cite{Bellucci:2006cx, Bellucci:2008uj, Lechner:2008uz}, especially in relation to the Bonora-Pasti-Tonin theorem \cite{Bonora:1986ix,Bonora:1987xn}.

\section*{Acknowledgments}
I would like to thank Eoin Colgáin, Junichi Sakamoto and Kentaroh Yoshida for helpful comments on the manuscript, as well as Arkady Tseytlin for an illuminating discussion on the non-locality of the counterterm.

\begin{appendix}

\section{Gamma matrices}
\label{gamma}

We recall the main properties of the $\gamma$-matrices, which are needed in the calculations.
In a Weyl representation, the $32\times 32$ matrices $\Gamma^a$ have the form
\begin{equation}
\label{Bia:Gamma}
	\Gamma^a = 
	\begin{pmatrix}
		0 & (\gamma^a)^{\alpha\beta} \\
		(\gamma^a)_{\alpha\beta} & 0 	
	\end{pmatrix}~,
\end{equation}
where the two sets of $16 \times 16$ matrices satisfy 
\begin{equation}
\label{Bia:Clifford}
	(\gamma^a)^{\alpha\gamma} (\gamma^b)_{\gamma\beta} + (\gamma^b)^{\alpha\gamma} (\gamma^a)_{\gamma\beta} 
	= 2 \eta^{ab} \delta^\alpha_\beta~.
\end{equation}
For example, one can take $(\gamma^0)^{\alpha\beta} = \delta^{\alpha\beta}$, $(\gamma^0)_{\alpha\beta} = -\delta_{\alpha\beta}$, and, for $a>0$, $(\gamma^a)^{\alpha\beta}= (\gamma^a)_{\alpha\beta}$, the $16 \times 16$ matrices generating the 9-d Euclidean Clifford algebra. However, the explicit form is not needed.

The basic Fierz identity is 
\begin{equation}
\label{Bia:Fierz}
	(\gamma^a)_{(\alpha\beta} (\gamma_a)_{\gamma)\delta} = 0~.
\end{equation}
From \eqref{Bia:Fierz}, one easily obtains the further Fierz identities
\begin{align}
\label{Bia:Fierz2}
	(\gamma^a)_{(\alpha\beta} (\gamma_a{}^{b_1\ldots b_{2n}})_{\gamma)\delta} &= -2n 
	(\gamma^{[b_1})_{(\alpha\beta} (\gamma^{b_2\ldots b_{2n}]})_{\gamma)\delta}~, \\
	(\gamma^a)_{(\alpha\beta} (\gamma_a{}^{b_1\ldots b_{2n+1}})_{\gamma)}{}^\delta &= -(2n+1) 
	(\gamma^{[b_1})_{(\alpha\beta} (\gamma^{b_2\ldots b_{2n+1}]})_{\gamma)}{}^\delta~. 
\end{align}
$(\gamma^a)_{\alpha\beta}$ and $(\gamma^{abcde})_{\alpha\beta}$ are symmetric, $(\gamma^{abc})_{\alpha\beta}$ anti-symmetric. Together, they form a basis of $16\times16$ matrices. This basis is over-complete, because the matrices $\gamma^{abcde}$ are self-dual.

\end{appendix}
\linespread{1.1}
\selectfont
\bibliography{kappaSym}
\end{document}